\documentclass[aps,twocolumn,showpacs,preprintnumbers,amsmath,amssymb,superscriptaddress,floatfix,nofootinbib]{revtex4}

\usepackage{graphicx}
\usepackage{epsfig}
\usepackage{epstopdf}
\usepackage{hyperref}
\usepackage{amsmath}
\usepackage{amsfonts}
\usepackage{amssymb}

\begin{document}

\title{The $K \bar K \pi$ decay of the $f_1(1285)$ and its nature as a $K^* \bar K -cc$ molecule}
\date{\today}
\author{F. Aceti}
\affiliation{Departamento de F\'{\i}sica Te\'orica and IFIC, Centro
Mixto Universidad de Valencia-CSIC Institutos de Investigaci\'on de
Paterna, Aptdo. 22085, 46071 Valencia, Spain}

\author{Ju-Jun Xie} \email{xiejujun@impcas.ac.cn}
\affiliation{Institute of Modern Physics, Chinese Academy of
Sciences, Lanzhou 730000, China} \affiliation{State Key Laboratory
of Theoretical Physics, Institute of Theoretical Physics, Chinese
Academy of Sciences, Beijing 100190, China}
\affiliation{Departamento de F\'{\i}sica Te\'orica and IFIC, Centro
Mixto Universidad de Valencia-CSIC Institutos de Investigaci\'on de
Paterna, Aptdo. 22085, 46071 Valencia, Spain}

\author{E.~Oset}
\affiliation{Departamento de F\'{\i}sica Te\'orica and IFIC, Centro
Mixto Universidad de Valencia-CSIC Institutos de Investigaci\'on de
Paterna, Aptdo. 22085, 46071 Valencia, Spain}
\affiliation{Institute of Modern Physics, Chinese Academy of
Sciences, Lanzhou 730000, China}

\begin{abstract}

We investigate the decay of $f_1(1285) \to \pi K \bar K$ with the
assumption that the $f_1(1285)$ is dynamically generated from the
$K^* \bar{K} - cc$ interaction. In addition to the tree level
diagrams that proceed via $f_1(1285) \to  K^* \bar{K} - cc \to \pi K
\bar K$, we take into account also the final state interactions of
$K \bar K \to K \bar K$ and $\pi K \to \pi K$. The partial decay
width and mass distributions of $f_1(1285) \to \pi K \bar K$ are
evaluated. We get a value for the partial decay width which, within
errors, is in fair agreement with the experimental result. The
contribution from the tree level diagrams is dominant, but the final
state interactions have effects in the mass distributions. The
predicted mass distributions are significantly different from phase
space and tied to the $K^* \bar{K} - cc$ nature of the $f_1(1285)$
state.

\end{abstract}

\maketitle
\raggedbottom

\section{Introduction}

The interaction of pseudoscalar mesons with vector mesons can be
tackled with the use of chiral Lagrangians \cite{birse}. These
chiral Lagrangians are also obtained by using the local hidden gauge
approach \cite{hidden1,hidden2,hidden3,hidden4}, exchanging vector
mesons between the vectors and the pseudoscalars in the limit of
small momentum transfers. Interesting developments using these
Lagrangians within a unitary scheme in coupled channels led to the
generation of the low lying axial vectors from the interaction of
these mesons, which qualify then as dynamically generated states
\cite{lutz,roca,thomas,valery,gengjuan}. The states could qualify as
kind of molecular states of a pair of mesons, or at least one can
claim that this is the dominant component in the wave function. One
of these resonances, the $a_1(1260)$ has been further investigated
and found to require some extra components, presumably $q \bar q$,
to explain some decay properties \cite{hideko}.  The extrapolation
of these ideas to the charm sector has also produced new states
\cite{Guo:2006rp,daniaxial,Guo:2009ct,Altenbuchinger:2013vwa} as the
$D_{s1}^*(2460)$, generated from $K D^*$. QCD lattice simulations
produce this latter state by using $K D^*$ interpolators
\cite{sasa}, suggesting a molecular nature for this resonance. A
more quantitative study has been done in Ref.~\cite{sasalberto} and
within errors of about 25 \% one determines in about 60 \% the
amount of $K D^*$ component in the wave function of that resonance.
The molecular nature of some resonances is catching interest, since
the structure is different than the standard $q \bar q$ commonly
accepted for mesons, and the recent developments with QCD lattice
simulations have revived this topic.

One of the cleanest example of these resonances is the $f_1(1285)$
with quantum numbers $I^G(J^P)=0^+(1^{++})$. This resonance appears
very clean and precise in Ref.~\cite{roca} from the single channel
$K^* \bar K -cc$, and the width is very small, as in the experiment,
because it cannot decay into two pseudoscalar mesons (in principle
$K \bar K$ in this case) for parity and angular momentum
conservation reasons. An extension of the work of Ref.~\cite{roca},
including higher order terms in the Lagrangian, has shown that the
effect of the higher order terms is negligible \cite{genghigher}.
Using these theoretical tools, predictions for lattice simulations
in finite volume have been done in Ref.~\cite{gengfinite}.

The width of the $f_1(1285)$ is $24 $ MeV, quite small for its mass,
and naturally explained within the molecular picture. Then the
channels contributing to it are very peculiar. For instance, the
$\pi a_0(980)$  channel accounts for 36\% of the width.  This
channel has been very well reproduced in Ref.~\cite{jorgifran}
within this molecular picture for the $f_1(1285)$, together with a
similar description of the $a_0(980)$ in the chiral unitary approach
from the interaction of pseudoscalar
mesons~\cite{npa,ramonet,kaiser,markushin,juanito,rios}. In
Ref.~\cite{jorgifran} the $\pi f_0(980)$ decay of the $f_1(1285)$
was also studied, and the rate and shape of the $\pi^+ \pi^-$ mass
distribution were predicted. These predictions have been confirmed
in a recent BESIII experiment~\cite{Ablikim:2015cob}.

Earlier work on the scalar resonances started from seeds of $q \bar
q$, which, after unitarization with the meson meson channels, gives
room to these meson meson channels which become dominant in the wave
function~\cite{van Beveren:1986ea,Tornqvist:1995ay,amir,amirscatt}.

On the other hand, there is another large channel,the $K \bar K
\pi$, which also accounts for about 9\% of the width.  This decay
channel should be tied to the $K \bar K^* -cc$ nature of the state.
The channel $K \bar K^*$ is bound for the energy of the $f_1(1285)$
by about 100 MeV, hence this decay is not observed
experimentally~\cite{pdg}. However, the decay of the $K^*$ off shell
can produce the $K \pi$ and then one has $K \bar K \pi$ in the final
decay channel. Definitely, this decay channel is related to the
coupling of the $f_1(1285)$ to the $K \bar K^* -cc$, and
consequently to the nature of this state.  Our aim in this paper is
to evaluate this decay channel from this perspective. In doing so we
also have to face the final state interaction (FSI) of the $K \bar
K$ and the $\pi K$, which we do using the chiral unitary
approach~\cite{npa,ramonet,dani}.

Apart from the tree level contribution, the FSI leads to loops with
one vector meson and two pseudoscalars. This triangular mechanism
was shown to be very important in the decay of the $\eta(1405)$ to
$\pi a_0(980)$ \cite{BESIII:2012aa} and the mixing with the isospin
violated $\pi f_0(980)$ channel~\cite{wuzhou,liangaceti}~\footnote{A
recent paper reviews this issue and, based on the contribution of
the imaginary part of the loop, concludes that there is a reduction
of the decay to the $\pi f_0(980)$ channel if the width of the $K^
*$ is considered \cite{achasov}. We have redone the calculations
including also the real parts and the reduction persists but is
weaker. However, the isospin allowed $\pi a_0(980)$ is more stable
and in the present case where we have a binding of the $K \bar K^*$
by 100 MeV the effect of the $K^ *$ width in the isospin allowed
channels is negligible.}. We follow the approach of
Refs.~\cite{wuzhou,liangaceti,jorgifran} to complement the tree
level contribution with the final state interaction of two mesons.
We show that the tree level contribution produces a decay rate of
the $f_1(1285)$ to $K \bar K \pi$ of the right order of magnitude,
while the final state interaction of two mesons is needed for a more
refined result, in good agreement with the experiment, hence
supporting the molecular nature of the $f_1(1285)$ resonance.

The picture that we present for the $f_1(1285)$ is somewhat
unconventional, and hence the need to find support for it, or
otherwise. The current trend up to now was that this resonance is a
simple $q \bar q$ state
\cite{xliu,klempt,gavillet,demingli,vijande,godfrey}. In
Ref.~\cite{xliu} the quark pair creation model is used to account
for decays of this resonance in two mesons and the $\pi a_0(980)$
decay is addressed from this perspective. In
Refs.~\cite{klempt,gavillet} the $f_1(1285)$ is assumed to belong to
a nonet of $q \bar q$ mesons. In Ref.~\cite{sheldons} the $B^0$ and
$B^0_s$ decays into $J/\psi$ and $f_1(1285)$ are investigated and
the results are interpreted in terms of a $q \bar q$ state, mostly
made of $u$ and $d$ quarks. Yet, in none of the works quoted, or
others, have we found an evaluation of the decay of this resonance
into $K \bar K \pi$.

\section{Formalism}

\begin{figure*}[htbp]
\begin{center}
\includegraphics[scale=0.5]{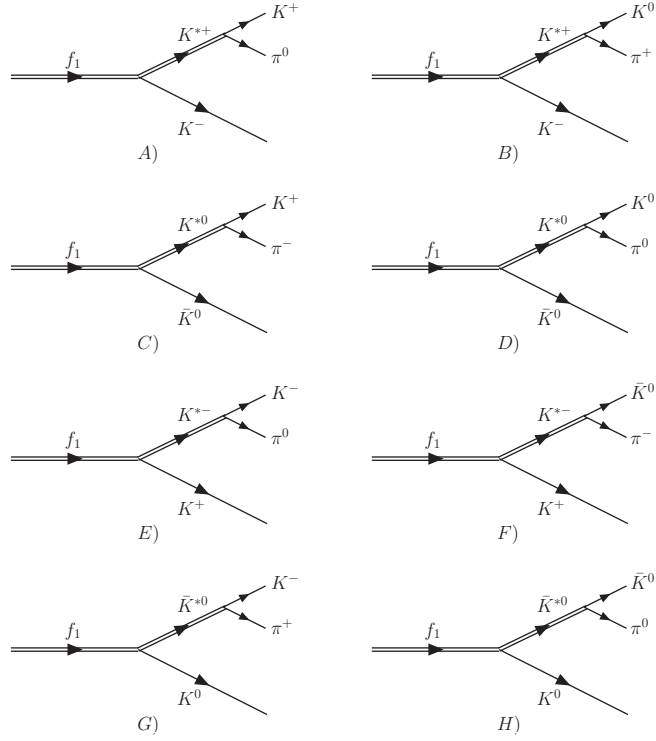}
\caption{Tree level diagrams representing the process $f_1(1285) \to
\pi K \bar{K}$.} \label{Fig:diagramTREE}
\end{center}
\end{figure*}

We study the decay of $f_1(1285) \to \pi K \bar{K}$ with the
assumption that the $f_1(1285)$ is dynamically generated from the
$K^* \bar{K}  - cc$ interaction, thus this decay can proceed via
$f_1(1285) \to  K^* \bar{K} - cc \to \pi K \bar K$. The tree level
diagrams are shown in Fig.~\ref{Fig:diagramTREE}.

\subsection{Decay amplitude at tree level}

In order to evaluate the partial decay width of $f_1(1285) \to \pi K
\bar{K}$, we need the decay amplitudes of the tree level diagrams
shown in Fig.~\ref{Fig:diagramTREE}, where the process is described
as the $f_1(1285)$ decaying to $K^* \bar{K} - cc $ and then the
$K^*$ decaying into $K \pi$. As mentioned above, the $f_1(1285)$
results as dynamically generated from the interaction of $K^*
\bar{K} - c.c.$. We can write the $f_1(1285) K^* \bar{K}$ vertex as
\begin{eqnarray}
\label{eq:vertex1}
-i t_1 = -i g_{f_1} C_1 \epsilon^{\mu} \epsilon'_{\mu},
\end{eqnarray}
where $\epsilon$ is the polarization vector of the $f_1(1285)$ state
and $\epsilon'$ is the polarization vector of the $K^*$
($\bar{K}^*$). The $g_{f_1}$ is the coupling constant of the
$f_1(1285)$ to the $ K^* \bar{K} -cc$ channel and can be obtained
from the residue in the pole of the scattering amplitude for
$\bar{K} K^* - c.c.$ in $I=0$. We take $g_{f_1} = 7555$ MeV in the
present calculation as it comes when the pole of the $f_1$ is made
to appear at the nominal mass of the $f_1(1285)$ resonance. This
result is in line but a bit bigger than the value of $7230$ MeV
found in Ref.~\cite{roca}, where a global fit to the axial vectors
was conducted.~\footnote{In Ref.~\cite{jorgifran} a coupling of
$g_{f_1} = 9687$ MeV was used, but this was based on an incorrect
method to evaluate the coupling. We take advantage here to say what
we would get with $g_{f_1} = 7555$ MeV. We obtain $BR(f_1(1285) \to
a_0(980) \pi)|_{\rm th} = (19 \pm 2)\%$, where we have added some
uncertainty induced by the discussion in the present work. This
should be compared with the experimental value of $BR(f_1(1285) \to
a_0(980) \pi) = (36 \pm 7)\%$. The agreement is, thus, at a
qualitative level.} We shall take the results with these two
couplings as a measure of the theoretical uncertainties. Besides,
the factors $C_1$ account for the weight of each $K^* \bar{K}$
($\bar{K}^* K$) component in the $I = 0$ and $C = +$ combination of
$K^* \bar{K}$ mesons, which is represented by

\begin{eqnarray}
\frac{1}{\sqrt{2}} (K^* \bar{K} - \bar{K}^* K) & = & - \frac{1}{2}
(K^{*+} K^- + K^{*0}\bar{K}^0 \nonumber \\
&& - K^{*-}K^+ - \bar{K}^{*0} K^0)\ .
\end{eqnarray}
We take the convention $CK^* = - \bar{K}^*$, which is
consistent with the standard chiral Lagrangians. Then we can easily
obtain the factors $C_1$ for each diagram shown in
Fig.~\ref{Fig:diagramTREE},
\begin{eqnarray}
C^{A,B}_1 = - \frac{1}{2};\ \ C^{C,D}_1 = - \frac{1}{2};\ \ C^{E,F}_1 = \frac{1}{2};\ \
C^{G,H}_1 = \frac{1}{2}\ .
\end{eqnarray}

To compute the decay amplitude, we also need the structure of the
$K^* K \pi$ vertices which can be derived using the hidden gauge
symmetry Lagrangian describing the vector-pseudoscalar-pseudoscalar
($VPP$) interaction~\cite{hidden1,hidden2,hidden3,hidden4}, given by

\begin{eqnarray}
{\cal L}_{VPP} = - i g <V^{\mu}[P,\partial_{\mu}P]>\ ,
\label{Eq:lvpp}
\end{eqnarray}
where $g = \frac{m_V}{2f}$ with $m_V \approx m_{\rho}$ and $f = 93$
MeV the pion decay constant. The symbol $<>$ stands for the
trace in $SU(3)$, while the $P$ and $V$ matrices contain the nonet of
pseudoscalar and vector mesons, respectively.

From the Lagrangian of Eq.~(\ref{Eq:lvpp}), the vertex of $K^* K
\pi$ can be written as

\begin{eqnarray}
\label{eq:vertex2}
-i t_2 = i g C_2 (k - p)^{\mu} \epsilon'_{\mu},
\end{eqnarray}
where $k$ and $p$ are the momenta of $\pi$ and $K$ mesons,
respectively. From Eq.~(\ref{Eq:lvpp}) and from the explicit expressions
of the $P$ and $V$ matrices, the factors $C_2$ for each diagram
shown in Fig.~\ref{Fig:diagramTREE} can be obtained,

\begin{eqnarray}
C^{A,H}_2 = \frac{1}{\sqrt{2}};\ \ C^{B,C}_2 = 1;\ \ C^{D,E}_2 =- \frac{1}{\sqrt{2}};\ \
C^{F,G}_2 = -1.
\end{eqnarray}

We can now sum the amplitudes of the diagrams that have same final state. By means
of Eqs. \eqref{eq:vertex1} and \eqref{eq:vertex2} and taking into account the
values of $C_1$ and $C_2$, the decay amplitude is obtained straightforwardly:

\begin{equation}
\label{eq:amplitudes}
\begin{split}
&M_{\rm tree}^{A+E}=M_{\rm tree}^{D+H}=M_{\rm tree}\ ,\\
&M_{\rm tree}^{B+G}=M_{\rm tree}^{C+F}=\sqrt{2}M_{\rm tree}\ ,
\end{split}
\end{equation}
with
\begin{eqnarray}
M_{\rm tree} &=& \frac{g g_{f_1}}{2\sqrt{2}} \big ( [-(k - p)_{\mu}
+ \frac{m^2_{\pi} -
m^2_K}{m^2_{K^*}}(k + p)_{\mu} ] D_1 \nonumber \\
&+& [-(k - p')_{\mu} +  \frac{m^2_{\pi} - m^2_K}{m^2_{K^*}}(k +
p')_{\mu}]D_2 \big ) \epsilon^{\mu} \nonumber \\
&=& \frac{g g_{f_1}}{2\sqrt{2}} \big ( [(\vec{k} - \vec{p}) -
\frac{m^2_{\pi} -
m^2_K}{m^2_{K^*}}(\vec{k} + \vec{p}) ] D_1\nonumber \\
&+& [(\vec{k} - \vec{p'}) + \frac{m^2_{\pi} -
m^2_K}{m^2_{K^*}}(\vec{k} + \vec{p'})] D_2 \big ) \cdot \vec
\epsilon ,
\end{eqnarray}
where
\begin{eqnarray}
D_1 &=& \frac{1}{(k + p)^2 - m^2_{K^*} + i m_{K^*} \Gamma_{K^*}}, \\
D_2 &=& \frac{1}{(k + p')^2 - m^2_{K^*} + i m_{K^*} \Gamma_{K^*}}\ .
\end{eqnarray}
Taking diagrams A) and E) for reference to calculate $M_{\rm tree}$, the
variables $p$, $p^{\prime}$ and $k$ refer to the $K^+$, $K^-$ and $\pi^0$,
and $\Gamma_{K^*}$ is the total decay width of the $K^*$ meson.

Since the dominant decay
channel of $K^*$ is $K \pi$, we can take
\begin{eqnarray}
 \Gamma_{K^*} = \Gamma_{\rm on} \left(\frac{q_{\rm on}}{q_{\rm off}}\right)^3 ,
\end{eqnarray}
with $\Gamma_{\rm on} = 49.1$ MeV, and
\begin{eqnarray}
q_{\rm on} &=& \frac{\lambda^{1/2}
(M^2_{K^*}, m^2_K, m^2_{\pi})}{2M_{K^*}} ,\\
q_{\rm off} &=& \frac{\lambda^{1/2} (M^2_{\rm inv}, m^2_K,
m^2_{\pi})}{2M_{\rm inv}} \theta(M_{\rm inv} \! - m_{K} \! -
m_{\pi}) ,
\end{eqnarray}
where $\lambda$ is the K\"allen function, with $\lambda(x, y, z) = (x
- y - z)^2 - 4yz$, and $M_{\rm inv}$ is the invariant mass of the
$\pi K$ system, which is $\sqrt{(k + p)^2}$ for the $D_1$ propagator and $\sqrt{(k
+ p')^2}$ for $D_2$.

\subsection{Decay amplitude for the triangular loop}

\begin{figure*}[htbp]
\begin{center}
\includegraphics[scale=0.6]{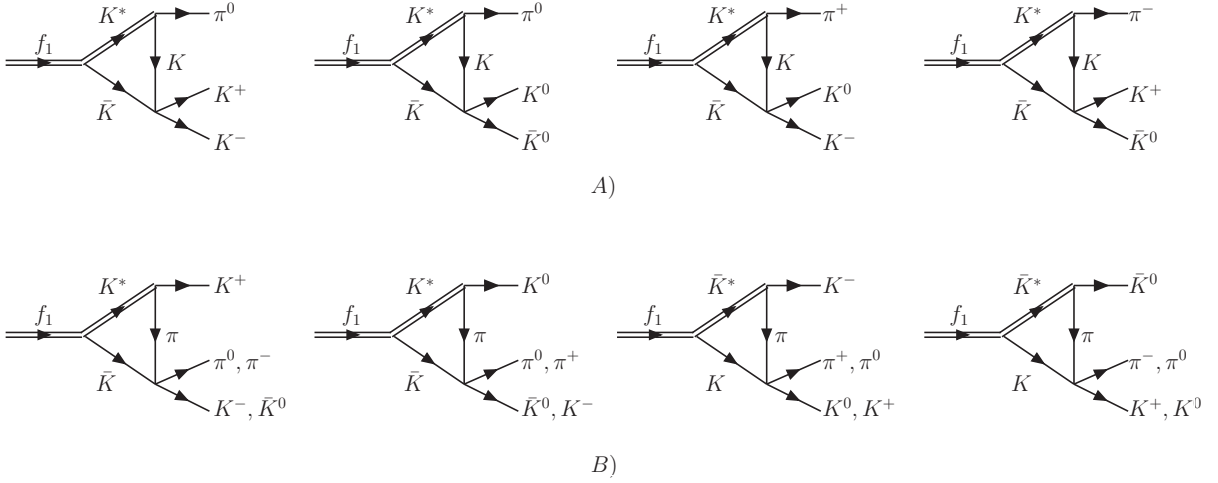}
\caption{Triangular loop contributions to the $f_1(1285) \to \pi
K \bar{K}$ decay.} \label{Fig:diagramFSI}
\end{center}
\end{figure*}

In addition to the tree level diagrams shown in
Fig.~\ref{Fig:diagramTREE}, we study also the contributions of the
$K \bar{K}$ and $\pi K$ FSIs. We use the triangular mechanism
contained in the diagrams shown in Fig.~\ref{Fig:diagramFSI},
consisting in the rescattering of the $K \bar{K}$ and $\pi K$ pairs.
Since the $f_1(1285)$ has $I=0$, considering only isospin conserving
terms, the $K\bar{K}$ will be in $I=1$ and the $\pi K$ in $I=1/2$.
The rescattering of the $K \bar{K}$ and $\pi K$ pairs with this
isospin dynamically generates in coupled channels the $a_0(980)$ and
$\kappa(800)$ resonances, respectively. We write for simplicity the
$K \bar K \to K \bar K$ and $\pi K \to \pi K$ rescattering
amplitudes as,
\begin{eqnarray}
t^{K \bar K}_{\rm FSI} (M_{K \bar K}) &= & t^{I = 1}_{K \bar K \to K \bar K} (M_{K \bar K}) , \\
t^{\pi K}_{\rm FSI} (M_{\pi K}) &= & t^{I = 1/2}_{\pi K \to \pi K}
(M_{\pi K}),
\end{eqnarray}
where $M_{K \bar K}$ and $M_{\pi K}$ are the invariant masses for
the $K \bar K$ and $\pi K$ systems, respectively. The quantities $t^{I = 1}_{K
\bar K \to K \bar K}$ and $t^{I = 1/2}_{\pi K \to \pi K}$ stand for
the scattering amplitudes of $K \bar K \to K \bar K$ in $I=1$ and
$\pi K \to \pi K$ in $I = 1/2$, respectively, and they can be obtained
using the Bethe-Salpeter equation
\begin{eqnarray}
t = (1 - VG)^{-1}V,
\end{eqnarray}
with the potential $V$ taken from Ref.~\cite{npa}. The $G$ function
in the above equation is the loop function for the propagators of
the intermediate particles
\begin{eqnarray}
G(P^2) = \int \frac{d^4q}{(2\pi)^4} \frac{1}{q^2 - m^2_1 + i
\epsilon} \frac{1}{(P - q)^2 - m^2_2 + i \epsilon},
\end{eqnarray}
where $P$ is the total four-momentum ($s = P^2$ is the invariant
mass square of the two particles in the loop) and $m_1$, $m_2$ the
masses of the particles in the considered channel. We take $K \bar
K$ and $\pi \eta$ channels for the case of $K \bar K$ FSI, while for
$\pi K$ FSI, we take $\pi K$ and $\eta K$ channels. After the
regularization by means of a cutoff~\cite{npa}, we obtain
\begin{eqnarray}
G(s) = \int_{|\vec{q}| < q_{\rm max}} \frac{d^3q}{(2\pi)^3}
\frac{\omega_1 + \omega_2}{2 \omega_1 \omega_2} \frac{1}{s -
(\omega_1 + \omega_2)^2 + i \epsilon} ,
\end{eqnarray}
with $\omega_i = \sqrt{|\vec{q}|^2 + m^2_i}$. For a good description
of $a_0(980)$ and $\kappa(800)$ we take a cutoff $q_{\rm max} = 900$
MeV, for both $K \bar K$ and $\pi K$ FSIs.

With the ingredients given above, we can explicitly write the decay
amplitude for the diagrams in Fig.~\ref{Fig:diagramFSI}. As for the
tree level case, we sum the diagrams with the same final state. In
Fig.~\ref{Fig:diagramFSI} A), we show the four possible final states
for the $K\bar{K}$ FSI. The amplitude corresponding to the first
diagram, that is the $\pi^0 K^+ K^-$ final state, is then given by
\begin{equation}
M^{K \bar K}_{\rm FSI} =  - \frac{g g_{f_1}}{2\sqrt{2}}
(2I_1 + I_2) 2 t^{I = 1}_{K \bar K \to K \bar K} (M_{K \bar K}) \vec \epsilon \cdot \vec k,
\label{Eq:tkakafsi}
\end{equation}
with $M_{K \bar K} = \sqrt{(p + p')^2}$. Here we have summed
explicitly the contributions of four diagrams corresponding to the
intermediate states $K^*K\bar{K}$: $K^{*+}K^-K^+$, $K^{*0}\bar{K}^0
K^0$, $K^{*-}K^+K^-$ and $\bar{K}^{*0}K^0\bar{K}^0$, easily done
taking into account the $C_1$ and $C_2$ coefficients and the fact
that
\begin{eqnarray}
t^{I=1}_{K\bar{K}\to K\bar{K}} &=& t_{K^+K^- \to K^+K^-} - t_{K^+K^- \to K^0 \bar{K}^0} \nonumber \\
&=& t_{K^0\bar{K}^0 \to K^0 \bar{K}^0} - t_{K^+K^- \to K^0\bar{K}^0}
,
\end{eqnarray}
with the phase convention $|K^-\rangle=-|1/2,-1/2\rangle$. The
quantities $I_1$ and $I_2$ for the case of $M^{K \bar K}_{\rm FSI}$
are given by
\begin{widetext}
\begin{eqnarray}
I_1 &=& - \int \frac{d^3q}{(2\pi)^3} \frac{1}{8 \omega(q)
\omega^{\prime}(q) \omega^*(q)} \frac{1}{k^0 - \omega^{\prime}(q) -
\omega^{*}(q) + i \epsilon} \frac{1}{P^0 - \omega^*(q) - \omega(q) + i \epsilon} \nonumber \\
&& \times \frac{2P^0 \omega(q) + 2k^0 \omega^{\prime}(q) -
2(\omega(q) + \omega^{\prime}(q))(\omega(q)+\omega^{\prime}(q) +
\omega^{*}(q))}{(P^0 -
\omega(q)-\omega^{\prime}(q)-k^0+i\epsilon)(P^0+\omega(q)+\omega^{\prime}(q)-k^0-i\epsilon)}  ,\label{Eq:loopintegral1}  \\
I_2 &=& -\int \frac{d^3q}{(2\pi)^3}
\frac{\vec{k}\cdot\vec{q}/|\vec{k}|^2}{8 \omega(q)
\omega^{\prime}(q) \omega^*(q)}
\frac{1}{k^0-\omega^{\prime}(q)-\omega^{*}(q) + i \epsilon}\,\frac{1}{P^0 - \omega^*(q) - \omega(q) + i \epsilon} \nonumber \\
&& \times \frac{2P^0 \omega(q) + 2k^0 \omega^{\prime}(q)-2(\omega(q)
+ \omega^{\prime}(q))(\omega(q) + \omega^{\prime}(q) +
\omega^{*}(q))}{(P^0-\omega(q) - \omega^{\prime}(q) -k^0 + i
\epsilon)(P^0 + \omega(q) + \omega^{\prime}(q) -k^0 -i \epsilon)}
,\label{Eq:loopintegral2}
\end{eqnarray}
\end{widetext}
where $\omega(q)  = \sqrt{\vec{q}^{~2} + m^2_K}$, $\omega'(q) =
\sqrt{(\vec q + \vec k)^2 + m^2_K}$, and $\omega^*(q) =
\sqrt{\vec{q}^{~2} + m^2_{K^*}}$ are the energies of the $K$
($\bar{K}$), $\bar K$ ($K$), and $K^*$ in the triangular loop,
respectively. A more detailed derivation can be found in Ref.
\cite{jorgifran}.

It is worth mentioning that after performing the integrations,
the $I_1$ and $I_2$ integrals in the above equations depend only on the
modulus of the momentum of the $\pi^0$, which can be easily related
to the invariant mass of the $K \bar K$ system via $M^2_{K \bar K} =
M^2_{f_1} + m^2_{\pi} - 2M_{f_1}\sqrt{|\vec{k}|^2 + m^2_{\pi}}$.
The $d^3q$ integrations are done with a cutoff $q_{\rm max} = 900$ MeV.

In the group B) of diagrams in Fig. \ref{Fig:diagramFSI}, we show
the possible final states corresponding to the $\pi K$ FSI. Each one
of the diagrams has two possible $\pi\bar{K}$ or $\pi K$ final
states. In addition, each one of the diagrams has two possible
$K^*\bar{K}$ or $\bar{K}^*K$ intermediate states: in the first
diagram we can have $K^{*+}K^-$ or $K^{*0}\bar{K}^0$ and this leads,
after considering the $C_1$ and $C_2$ coefficients to the
combination
$t_{\pi^0K^-\rightarrow\pi^0K^-}+\sqrt{2}t_{\pi^-K^0\rightarrow\pi^0K^-}$,
proportional to the $t^{I=1/2}_{\pi K\rightarrow\pi K}$. The sum of
the first and third diagram with $\pi^0K^+K^-$ in the final state is
then easily done and can be cast as
\begin{equation}
\begin{split}
M^{\pi K}_{\rm FSI} &= \frac{g g_{f_1}}{2\sqrt{2}} (2I_1^{\prime} + I_2^{\prime})
t^{I = 1/2}_{\pi K \to \pi K} (M^{(1)}_{\pi K}) \vec \epsilon \cdot
\vec p\\
&+ \frac{g g_{f_1}}{2\sqrt{2}} (2I_1^{\prime\prime} + I_2^{\prime\prime}) t^{I = 1/2}_{\pi K \to
\pi K} (M^{(2)}_{\pi K}) \vec \epsilon \cdot \vec{p'} , \label{Eq:tpikafsi}
\end{split}
\end{equation}
where now $I_1^{\prime}$, $I_2^{\prime}$ are evaluated with Eqs.
\eqref{Eq:loopintegral1} and \eqref{Eq:loopintegral2} replacing one
kaon propagator by a pion and simply putting
$\omega^{\prime}(q)=\sqrt{(\vec{q}+\vec{p})^2+m_{\pi}^2}$ and
substituting $k^0$ by $p^0$. Similarly $I_1^{\prime\prime}$ and
$I_2^{\prime\prime}$ are also evaluated with Eqs.
\eqref{Eq:loopintegral1} and \eqref{Eq:loopintegral2} putting
$\omega^{\prime}(q)=\sqrt{(\vec{q}+\vec{p}^{\ \prime})^2+m_{\pi}^2}$
and substituting $k^0$ by $p^{\prime 0}$. The integrals
$I_1^{\prime}$, $I_2^{\prime}$ are functions of $|\vec{p}|$ and
$I_1^{\prime\prime}$, $I_2^{\prime\prime}$ of $|\vec{p}^{\
\prime}|$, which can be written in terms of the invariant masses
$M_{\pi K}^{(1)}=\sqrt{(k+p^{\prime})^2}$ and $M_{\pi
K}^{(2)}=\sqrt{(k+p)^2}$ respectively, similarly as done before for
the $K\bar{K}$ interaction terms.

The relative minus sign between Eqs.~(\ref{Eq:tkakafsi}) and (\ref{Eq:tpikafsi}) is easily
traced back to the sign of the $K^* \to K \pi$ when we have either
the $K$ or the $\pi$ in the loop.

\section{Numerical results}

With the decay amplitudes obtained above, we can easily get the
total decay width of $f_1(1285) \to \pi K \bar{K}$ which is
\begin{equation}
\begin{split}
\Gamma &= 6 \frac{1}{64 \pi^3 M_{f_1}} \int \int d\omega_{K^+}
d\omega_{K^-} \overline{\sum} |M|^2 \\ &\times\theta(1 - {\rm cos}^2\theta_{K \bar K})
\theta(M_{f_1} - \omega_{K^+} - \omega_{K^-} - m_{\pi}),
\label{eq:gamma1}
\end{split}
\end{equation}
where $M$ is the full amplitude of the process $f_1(1285)\rightarrow\pi^0K^+K^-$
including the FSIs,
\begin{equation}
\label{eq:fullampl}
M = M_{\rm tree} + M^{K \bar K}_{\rm FSI} + M^{\pi K}_{\rm FSI},
\end{equation}
with $M_{f_1} = 1281.9$ MeV the mass of $f_1(1285)$ state and
$\omega_{K^+} = \sqrt{m^2_K + \vec{p}^{~2}}$ and $\omega_{K^-} =
\sqrt{m^2_K + \vec{p'}^{~2}}$ the energies of the $K^+$ and $K^-$
mesons, respectively. The symbol $\overline{\sum}$ stands for the
average over the polarizations of the initial $f_1(1285)$ state. The
factor 6 in the formula of $\Gamma$ accounts for the different final
charges for $\pi K \bar K$: $\pi^0 K ^+ K^-$, $\pi^+ K^0 K^-$,
$\pi^- K^+ \bar{K}^0$, and $\pi^0 K^0 \bar{K}^0$, having weights 1,
2, 2, and 1, respectively, which can be easily obtained using simple
Clebsch-Gordan coefficients. Besides, the ${\rm cos}\theta_{K \bar
K}$ is defined by energy conservation as

\begin{eqnarray}
{\rm cos}\theta_{K \bar K} &=& \frac{1}{2|\vec{p}| |\vec{p}^{\ \prime}|}
[M^2_{f_1} + 2m^2_K - 2M_{f_1} (\omega_{K^+} + \omega_{K^-})
\nonumber \\
&& + 2 \omega_{K^+} \omega_{K^-} - m^2_{\pi}].
\end{eqnarray}

With the full amplitude of Eq.~\eqref{eq:fullampl}, the numerical
result for the partial decay width is, using $g_{f1} = 7555$ MeV,
$\Gamma=1.9$ MeV, which corresponds to a branching ratio
\begin{eqnarray}
B.R.[f_1(1285) \to \pi K \bar K] = 7.8 \%.
\end{eqnarray}
If we use the coupling of Ref.~\cite{roca}, $g_{f1}=7230$ MeV, then
we get $\Gamma=1.74$ MeV, corresponding to a branching ratio
\begin{eqnarray}
B.R.[f_1(1285) \to \pi K \bar K] = 7.2 \%.
\end{eqnarray}
This gives a band of theoretical results of
\begin{eqnarray}
B.R.[f_1(1285) \to \pi K \bar K] = (7.2 - 7.8) \%,
\label{Eq:bruncertainty}
\end{eqnarray}
which is in fair agreement with the experimental value: $(9.0  \pm
0.4) \%$~\cite{pdg,Barberis:1997vf,Barberis:1998by}. The result
would be $9.1\%$, with the big $g_{f1}$ coupling, if we considered
only the tree level diagrams. This indicates that the contribution
from the FSIs is small. This occurs because of the relative minus
sign in Eqs.~(\ref{Eq:tkakafsi}) and (\ref{Eq:tpikafsi}), which
makes the effects of the FSIs for $K \bar K$ and $\pi K$ go in
opposite directions bringing a partial cancelation in $\Gamma$.

We should take into account that in order to get the $f_1(1285)$
state, cut offs of the order of $1000$ MeV in the $G$ function of
$\bar{K} K^*$ are used. On the other hand for the $G$ function of
$K\bar{K}$ and $\pi K$ a cut off of $900$ MeV was used. In the
triangular loop function of Fig.~\ref{Fig:diagramFSI} we have then
$\theta(1000 - |\vec{q}|)~\theta(900 - |\vec{q}|) = \theta(900 -
|\vec{q}|)$ (in MeV). This justifies the choice of $q_{\rm max}$ in
that loop function. We can see the variation of our results by
changing these cut offs in a range such that the masses of the
$f_1(1285)$ and $a_0(980)$ are not much changed with respect to the
experimental values. In this sense, changes of $q_{\rm max}$ from
$980$ MeV to 1040 MeV bring changes in the mass of the $f_1(1285)$
by $12$ MeV and only $1\%$ changes in the couplings. These changes
are smaller than the range of couplings accepted in
Eq.~(\ref{Eq:bruncertainty}). Similarly, changes in $q_{\rm max}$
for $a_0(980)$ from $860$ MeV to $940$ MeV change the mass of the
$a_0(980)$ in $7$ MeV. Reevaluating the branching ratios with values
of $q_{\rm max}$ within this range, change the result that we quote
in Eq.~(\ref{Eq:bruncertainty}) to
\begin{eqnarray}
B.R.[f_1(1285) \to \pi K \bar K] = (7.2 - 8.3) \%,
\end{eqnarray}
with the upper limit a little closer to the experimental value.

Next, we study the invariant mass distribution of the $f_1(1285) \to
\pi^0 K^+ K^-$ decay to see the effect of the $K^*$ propagator in
the tree level and of the $K \bar K$ and $\pi K$ FSIs.

The invariant mass distributions are given by the formulas
\begin{eqnarray}
\frac{d\Gamma}{dM_{K^+ K^-}} &=& \frac{M_{K^+ K^-}}{64 \pi^3
M^2_{f_1}} \int
d\omega_{K^+} \overline{\sum} |M|^2 \theta(1 - {\rm cos}^2\theta_{K \bar K}) \times \nonumber \\
&& \!\!\!\!\!
\theta(M_{f_1} - \omega_{K^+} - \omega_{K^-} - m_{\pi}) \theta(\omega_{K^-} - m_K), \label{Eq:dgdkaka} \\
\frac{d\Gamma}{dM_{ \pi^0 K^+}} &=& \frac{M_{\pi^0 K^+}}{64 \pi^3
M^2_{f_1}} \int
d\omega_{K^+} \overline{\sum} |M|^2 \theta(1 - {\rm cos}^2\theta_{K \bar K}) \times \nonumber \\
&& \!\!\!\!\! \theta(M_{f_1} - \omega_{K^+} - \omega_{K^-} -
m_{\pi}) \theta(\omega_{K^-} - m_K),  \label{Eq:dgdpika}
\end{eqnarray}
where
\begin{eqnarray}
\omega_{K^-} = \frac{1}{2M_{f_1}}(M^2_{K^+ K^-} + M^2_{f_1}
-m^2_{\pi}) - \omega_{K^+},
\end{eqnarray}
for $\frac{d\Gamma}{dM_{K^+ K^-}}$, while
\begin{eqnarray}
\omega_{K^-} = \frac{1}{2M_{f_1}}(M^2_{f_1} + m^2_K - M^2_{\pi^0
K^+}),
\end{eqnarray}
for $\frac{d\Gamma}{dM_{ \pi^0 K^+}}$.

\begin{figure}[htbp]
\begin{center}
\includegraphics[scale=0.45]{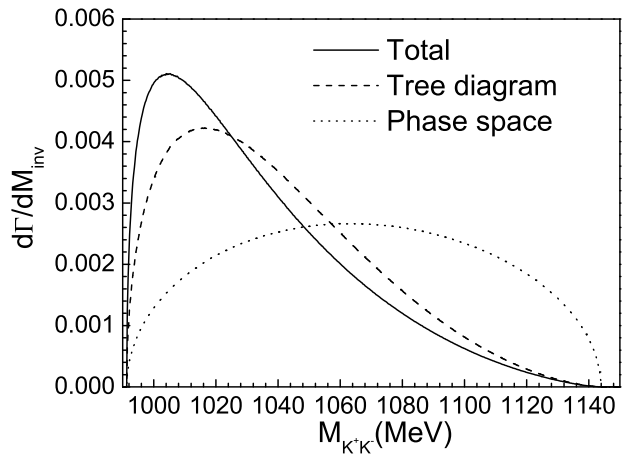}
\caption{The mass distribution $\frac{d\Gamma}{dM_{K^+ K^-}}$ for
$f_1(1285) \to \pi^0 K^+ K^-$ as a function of the invariant mass of
the $K^+ K^-$ system.} \label{Fig:dgdm-kaka}
\end{center}
\end{figure}

\begin{figure}[!ht]
\begin{center}
\includegraphics[scale=0.45]{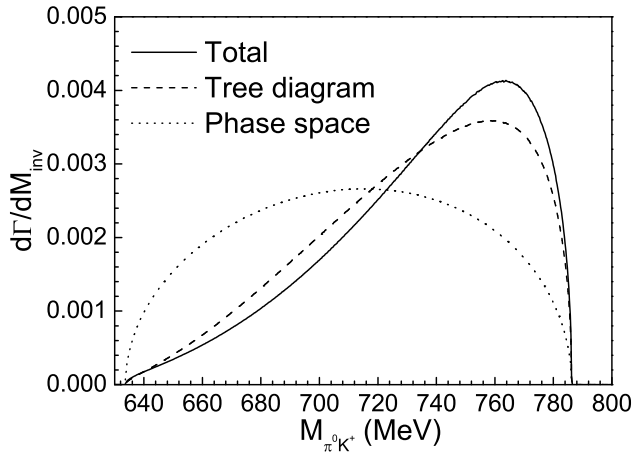}
\caption{The mass distribution $\frac{d\Gamma}{dM_{\pi^0 K^+}}$ for
$f_1(1285) \to \pi^0 K^+ K^-$ as a function of the invariant mass of
the $\pi^0 K^+$ system.} \label{Fig:dgdm-pika}
\end{center}
\end{figure}

The results for $\frac{d\Gamma}{dM_{K^+ K^-}}$ and
$\frac{d\Gamma}{dM_{ \pi^0 K^+}}$ are shown in
Fig.~\ref{Fig:dgdm-kaka} and Fig.~\ref{Fig:dgdm-pika}, respectively.
It is very interesting to compare the different curves in
Figs.~\ref{Fig:dgdm-kaka} and \ref{Fig:dgdm-pika}. We show there the
results assuming just a phase space distribution ($\overline{\sum}
|M|^2$ in Eqs.~(\ref{Eq:dgdkaka}) and (\ref{Eq:dgdpika}) is set to a
constant), and with the tree level or tree level plus final state
interaction of $K \bar K$ and $\pi K$. For the sake of comparison,
the curves are normalized to the same $\Gamma$. In
Fig.~\ref{Fig:dgdm-kaka} we see that the tree level alone shows a
distinct shape, very different from phase space, with a peak at low
$M_{K^+ K^-}$. This must be attributed to the effect of the $K^*$ off
shell propagator. The implementation of FSI, particularly the $K
\bar K$ in this case, is responsible for a further shift of the
mass distribution to lower invariant masses, closer to the $K \bar K$
threshold, where the $a_0(980)$ resonance appears.

In Fig.~\ref{Fig:dgdm-pika}, where the $\pi K$ invariant mass
distribution is plotted, we see a similar behaviour. The tree level
alone already produces a shape quite different from phase space,
with a peak at high values of $M_{\pi K}$, to be attributed once again to the off
shell $K^*$ propagator. The implementation of FSI, particularly the
$\pi K$ in this case, pushes the peak of the mass
distributions to higher $M_{\pi K}$, closer to the region where the
$\kappa(800)$ resonance appears.

The two figures show how the most drastic change in the shape of the
two mass distributions is already caused by the tree level alone
and, as mentioned before, this is tied to the $K^*$ propagators,
which appears at tree level because of the $\bar K^* K  -cc$ nature
of the $f_1(1285)$ state that we have assumed. These mass
distributions have not been measured yet and it is clear from the
present study that their observation would be very important to
determine the nature of this resonance.

So far we have assumed that the $f_1(1285)$ resonance is fully made
from $K \bar K^*$. There are hints that the resonance could have
also other components. Indeed, in the study of this resonance in
finite volume~\cite{gengfinite} it was shown that applying the
compositeness sum rule~\cite{hyodo1,hyodo2,sekihara} to this case
with the chiral potential, the $K \bar K^*$ molecular component
accounted for about $50\%$ of the probability of the wave function,
but there could still be a sizable fraction for other non $K \bar
K^*$ components. The size of these components is uncertain because
it relies on the energy dependence contained in the chiral potential
and it is unclear that this accounts for missing channels (see
Ref.~\cite{acetidelta}), but it really hints at the possibility to
have some non negligible non $K \bar K^*$ molecular component in the
$f_1(1285)$ wave function. This might seem to be in conflict with
our claims of a basically molecular state for this resonance. This
requires some explanation. Different parts of the wave function
revert in different ways on certain observables. The easiest such
case is the nucleon form factor, which at low momentum transfer is
dominated by the meson baryon components of the nucleon, while at
high momentum transfers it is the core of quarks that is responsible
for it~\cite{thomasrep,ulfrep}. In this sense, it is logical that
the decay of the $f_1(1285)$ into $K \bar K \pi$ and related
channels is mostly due to the $K \bar K^*$ molecular component of
the wave function, and other components would show up in other
reactions. In this sense it is interesting to note that in
Ref.~\cite{sheldons} the $B^0$ and $B^0_s$ decays into $J/\psi$ and
$f_1(1285)$ are investigated and the interpretation in terms of a $q
\bar q$ state leads to a $f_1(1285)$ state mostly made of $u$ and
$d$ quarks. In our case we have four quarks to start with and a
sizable fraction of strange quarks in our $K \bar K^*$ molecular
component, so the models seem to be contradictory. Yet, one must
recall that in this latter case we have production of the resonance
in $B$ decays and the resonance must be formed starting from a $q
\bar q$ component. The investigation done in
Refs.~\cite{liang,liangmela} of the $B$ decays, and the ratio of the
rates of $\bar{B}^0 \to J/\psi f_0(500)$~\cite{Aaij:2013zpt} and
$\bar{B}^0 \to J/\psi \rho$~\cite{pdg}, show that the hadronization
of the primary $q \bar q$ component to give two mesons has a penalty
factor that reverts into a factor of $0.37$ decrease in the partial
decay width. In this sense, the decays of heavy mesons leading to
light ones might reveal themselves into a source of information on
the non molecular components of states like the present one. Further
research considering both the molecular and $q\bar{q}$ components
for this resonance in the $B$ decays would be most welcome after the
discussion made here.

\section{Summary}

In this work, we evaluate the partial decay width of the $f_1(1285)
\to \pi K \bar K$ with the assumption that the $f_1(1285)$ is
dynamically generated from the $\bar K^* K - cc$ interaction. The
tree level diagrams proceeding via $f_1(1285) \to  K^* \bar{K} -cc
\to \pi K \bar K$ are considered. Besides, we also take into account
the final state interactions of $K \bar K \to K \bar K$ and $\pi K
\to \pi K$. It is found that the contributions from the FSIs are
small compared to the tree level diagrams to the partial decay
width, but they change the mass distributions of the $f_1(1285) \to
\pi K \bar K$ decay.

The result that we obtained for the width is compatible with
experiment within errors. Yet, we find some relevant features in the
$K \bar K$ and $\pi K$ mass distributions, which turn out to be very
different from phase space. The FSI is partly responsible for the
shapes obtained but we found that the tree level contribution, which
is dominant in the process, is mostly responsible for this different
shape, which must be attributed to the off shell $K^*$ propagator
appearing in the process under the assumption that the $f_1(1285)$
is a $K^* \bar{K} - cc$ molecule. The experimental observations of
those mass distributions would then provide very valuable
information on the relevance of this component in the $f_1(1285)$
wave function.

\section*{Acknowledgments}

One of us, E. O., wishes to acknowledge support from the Chinese
Academy of Science (CAS) in the Program of Visiting Professorship
for Senior International Scientists (Grant No. 2013T2J0012). This
work is partly supported by the Spanish Ministerio de Economia y
Competitividad and European FEDER funds under the contract number
FIS2011-28853-C02-01 and FIS2011-28853-C02-02, and the Generalitat
Valenciana in the program Prometeo II, 2014/068. We acknowledge the
support of the European Community-Research Infrastructure
Integrating Activity Study of Strongly Interacting Matter (acronym
HadronPhysics3, Grant Agreement n. 283286) under the Seventh
Framework Programme of EU. This work is also partly supported by the
National Natural Science Foundation of China under Grant No.
11475227.

\bibliographystyle{plain}

\end{document}